\documentclass[twocolumn,prd,showpacs,superscriptaddress,groupedaddress,preprintnumbers,nofootinbib]{revtex4-1}
\pdfoutput=1
\usepackage{graphicx,amsfonts,amsmath,amssymb,epsfig,color, mathrsfs,latexsym,url,wasysym,float,xfrac}
\newcommand{\be}{\begin{equation}}
\newcommand{\ee}{\end{equation}}
\newcommand{\nn}{\nonumber}
\newcommand{\ba}{\begin{eqnarray}}
\newcommand{\ea}{\end{eqnarray}}
\newcommand{\mpl}{m_{\rm Pl}}

\begin{document}

\preprint{SUT/Physics-nnn}
\title{5-Dimensional Chern-Simons Gauge Theory on an Interval:\\Massive Spin-2 Theory from Symmetry Breaking via Boundary Conditions}
\author{Mahdi Torabian$^*$}
\affiliation{Department of Physics, Sharif University of Technology, Azadi Ave, 1458889694, Tehran, Iran}

\begin{abstract}
In this note, we revisit the 4-dimensional theory of massive gravity through compactification of an extra dimension and geometric symmetry breaking. We dimensionally reduce the 5-dimensional topological Chern-Simons gauge theory of (anti) de Sitter group on an interval. We apply non-trivial boundary conditions at the endpoints to break all of the gauge symmetries. We identify different components of the gauge connection as invertible vierbein and spin-connection to interpret it as a gravitational theory. The effective field theory in four dimensions includes the dRGT potential terms and has a tower of Kaluza-Klein states without massless graviton in the spectrum. The UV cut of the theory is the Planck scale of the 5-dimensional gravity $l^{-1}$. If $\zeta$ is the scale of symmetry breaking and $L$ is the length of the interval, then the masses of the lightest graviton $m$ and the level $n$ (for $n<Ll^{-1}$) KK gravitons $m_{\rm KK}^{(n)}$ are determined as $m=(\zeta L^{-1})^{\sfrac{1}{2}}\ll m_{\rm KK}^{(n)}=nL^{-1}$. The 4-dimensional Planck mass is $m_{\rm Pl}\sim (Ll^{-3})^{\sfrac{1}{2}}$ and we find the hierarchy $\zeta< m< L^{-1}<l^{-1}<m_{\rm Pl}$.
\end{abstract}

\maketitle

\subsection*{I. Introduction} 
The conventional Lorentz covariant formulation of fields in tensorial representations demands special effort. The action terms and the coefficients must be wisely picked so that the redundant unphysical ghost-like degrees of freedom are not propagating. The highest helicity modes get propagation through the derivative terms and the lower ones from the non-derivative terms. A gauge symmetry is emerged as a result of physical consistency conditions which is useful to check consistencies, even if it is softly broken, throughout computations. 

The General Relativity is the unique consistent interacting theory of massless spin-2 particles. It is an effective field theory valid up to the 4-dimensional Planck scale $\mpl$. It is invariant under active diffeomorphisms of the dynamical metric through which only the transverse helicity-2 modes propagate. The longitudinal modes in a massive spin-2 particle get propagation by the potential terms. The Fierz-Pauli mass terms excite longitudinal modes yet, with a tuning of parameters, the ghost-like sixth mode is kept non-dynamical \cite{Fierz:1939ix}. It took a long time to construct a ghost-free non-linear completion of massive gravity which is known as the de Rham-Gabadadze-Tolley (dRGT) potential terms \cite{deRham:2010gu,deRham:2010ik,deRham:2010kj,Hassan:2011vm,Hassan:2011hr,deRham:2011rn,deRham:2011qq,Hassan:2011ea,Hassan:2012qv,Gabadadze:2013ria}. It is a 2-parameter family of effective field theories with a UV cutoff $\Lambda_3=(\mpl m^2)^{\sfrac{1}{3}}$ \cite{ArkaniHamed:2002sp,ArkaniHamed:2003vb,Schwartz:2003vj} . We note that it is proportional to the graviton mass $m$ which modifies gravity at long distances. For cosmologically interesting value of graviton mass of order the present Hubble rate $m\sim H_0\sim 10^{-33}$ eV we find $\Lambda_3\sim 10^{-13}$ eV. However, we expect that $\Lambda_{\rm UV}\gtrsim 10^{-3}$ eV from sub-millimeter tests of gravity which is parametrically close to the naive expectation for a would-be UV cutoff  $\Lambda_2=(\mpl m)^{\sfrac{1}{2}}$ (see \cite{Rubakov:2008nh,Hinterbichler:2011tt,deRham:2014zqa} for reviews on massive gravity).

The $\Lambda_3$ cutoff is computed via the unitarity bound from the tree-level scattering amplitudes before the theory becomes strongly coupled. In fact, the longitudinal modes pose a serous difficulty, as they are derivatively-coupled, when the scattering amplitudes are calculated. The amplitudes rapidly grow with (the center of) energy $s$ and quickly hit the unitarity bound. In the dRGT theory the amplitudes grow like $s^3$ and thus the theory is applicable below $\Lambda_3$. We expect that it is UV completed in a theory with more particles/interactions such that the cutoff is, parametrically close to $\mpl$. Although not yet achieved, there are extensive studies which show that massive spin-2 theory admit a perturbative local Lorentz invariant UV completion \cite{Cheung:2016yqr,Bellazzini:2016xrt,Bonifacio:2016wcb,deRham:2017imi,deRham:2017zjm,Bellazzini:2017fep,deRham:2017xox,Bonifacio:2018vzv,deRham:2018qqo,Bonifacio:2018aon,Alberte:2019zhd}. 

We recall that in massive spin-1 theory, the high energy behavior of a scattering amplitude involving longitudinal modes can be improved by the exchange of a scalar ({\it a.k.a.} the Higgs) field. The massive spin-1 is perturbatively UV completed and the masses and the couplings are determined through the mechanism of spontaneous symmetry breaking. We might imagine a similar scenario for massive spin-2 theory. Indeed in \cite{Torabian:2017bqu}, we attempted to build the dRGT theory as a result of spontaneous symmetry breaking (see \cite{Percacci:1990wy,Kirsch:2005st,tHooft:2007rwo,Chamseddine:2010ub,Alberte:2010it,Alberte:2010qb,Dubovsky:2004sg} for earlier studies of Higgs-like mechanism in gravity). We proposed a topological gauge theory with {\sl ISO(1,3)} gauge group (or a  limit of {\sl SO(1,4}) group) which could be interpreted as a gravitational theory in the space of invertible gauge connections. There is a symmetry breaking phase with residual diagonal {\sl SO(1,3)} global symmetry in the vacuum. The effective theory around this minimum is precisely the dRGT theory plus an additional interacting Higgs field. Although we succeeded to construct the  ghost-decoupling structure of the dRGT theory and determined its parameters from top-down, further analysis showed that the extra scalar mode cannot improve the UV behavior. Remarkably in  \cite{Bonifacio:2019mgk}, it was argued that the $\Lambda_3$ is the highest cutoff scale if no other massive spin-2 excitations are present. In the other words, in contrast to the spin-1 case, the exchange of any number of scalar and vector fields cannot ameliorate the rapid growth of the scattering amplitudes (See \cite{deRham:2016plk,Gabadadze:2017jom,Gabadadze:2019lld}
 for attempts to scale up the UV cutoff in different approaches).

On the other hand, the 4-dimensional spacetime might be a subspace of a higher-dimensional spacetime. Theories in higher dimensions provide a simpler (and often geometric) explanation of physics in lower dimensions. For instance, the Kaluza-Klein (KK) dimensional reduction offers a mechanism to build theories of interacting massive spin-1 and spin-2 particles with controllable scattering amplitudes without a Higgs excitations. As expected, it is shown that  the scattering amplitudes of longitudinal modes of massive spin-1 particles would not grow like $s$ \cite{Chivukula:2003kq}. Cancellation occurs by the exchange of massive spin-1 KK states and the unitarity is guaranteed by the presence of the entire KK tower. 
Indeed, this results from the gauge symmetry of the higher-dimensional theory. 
The UV cutoff of the lower-dimensional theory is that of the higher-dimensional gauge theory and we say that the theory with all KK states is UV completed in higher dimensions. Recently, it is argues that the high energy conduct of scattering amplitudes of massive spin-2 particles are similarly improved if the whole tower of KK modes are included \cite{Chivukula:2019rij,Bonifacio:2019ioc,Chivukula:2019zkt}. Dimensional reduction of General Relativity gives General Relativity in lower dimensions plus interacting massive modes whose coupling are dictated by the Einstein-Hilbert action in higher dimensions. The spectrum is composed of massless spin-2, spin-1 and spin-0 particles and their massive KK counterpart. The UV cutoff of the lower-dimensional gravity is the Planck scale in higher dimensions. 

Moreover, extra dimensions not only naturally introduces massive states by compactification, but also can be applied to break all or some of gauge symmetries in lower dimensions. For instance, if the Yang-Mills theory is compactified on an interval, non-trivial boundary conditions at endpoints can reduce the gauge symmetry in the lower-dimensional theory \cite{Hebecker:2001jb}. Interestingly, it was shown that this geometric symmetry breaking is soft and, given the entire KK modes, the scattering amplitude of longitudinal modes is well-behaved \cite{Csaki:2003dt}. The symmetry is broken by boundary conditions on gauge fields, the massive spin-1 appears in the spectrum by dimensional reduction and unitarity is preserved by the exchange of massive vectors.  The choice of boundary condition can be such that the gauge symmetry is completely broken in lower dimensions and no massless gauge field appears. The gauge symmetry is a useful concept to get control of propagation and interactions of physical modes. When it is softly broken (to give more interesting physics) its virtues is descended to the broken phase. 

Motivated by the above results, in this paper we construct a theory of massive spin-2 particles from compactification of an extra dimension and geometric symmetry breaking. We choose the boundary conditions at the endpoint of an interval such that there is no massless spin-2 mode in the spectrum. That makes the theory similar to the dRGT construction. Massive spin-2 KK states improving the high energy behavior of the scattering amplitudes. Besides the interaction terms induced by the Einstein-Hilbert action, there are interactions from the dRGT-like terms. However, as the symmetry breaking by these contributions are soft, we expect that the scattering amplitudes of 2-to-2 gravitons  are not divergent worse than $s$. 

It has long been known that the first-order formalism of gravitational theories in odd dimensions admit gauge theory formulations in terms of Chern-Simons (CS) theories \cite{Witten:1988hc,Witten:1989sx,Chamseddine:1989nu} (see  \cite{Morales:2016qrx,Albornoz:2018uin} for more recent studies). Different components of the gauge connection can be identified as geometrical quantities such the vielbeins and the spin-connection to define the metricity and affinity. Then, the $d$-dimensional diffeomorphism invariance {\sl Diff(d)} in the metric theory is descended from the topological structure of the gauge theory. 

In the present study,  we start from a 5-dimensional topological CS theory of (anti) de Sitter gauge group. Then to get a 4-dimensional theory, we compactify the extra dimension on an interval. By varying the action in the presence of endpoints, we find the bulk equations of motion and the boundary terms that must be simultaneously vanishing. As we will see, different choices for different components of gauge fields implies different physics in lower dimensions. After identification of geometrical connections, we obtain the General Relativity plus KK modes for one choice and the dRGT theory  with KK states for the other option. In high energy limit with all KK modes included, the 4 dimensional scattering amplitude must match that of 5 dimensional theory and perturbative unitarity is guaranteed from contributions from different KK states. Therefore, the UV cutoff of the effective theory in 4-dimensions is the 5-dimensional Planck scale (see also \cite{Gabadadze:2009ja,deRham:2009rm} where boundary conditions of a spurious extra dimension induce a mass term).
 
The rest of this paper is organized as follows.  In the next section we briefly review the CS gauge theory in five dimensions in different representations and perform partial gauge fixing. Then, we find the equations of motion, determined the boundary terms and present consistent boundary conditions. Then, we compatify the theory to four dimensions where we find, besides the Einstein-Cartan action, the dRGT potential terms in the first order formalism. We identify the parameters of massive gravity in terms of the fundamental parameters of the compactified CS theory. Next, we compute the coupling constants of interaction induced by the potential terms. Finally, we conclude the results in the last section. 
  
\subsection*{II. Chern-Simons theory in 5 dimensions}
We start by considering the CS topological gauge theory in 5 dimension. The gauge connection 1-form ${\bf A}={\textstyle\frac{1}{2}}A^{IJ}t_{IJ}$ is valued in {\sl so(1,5)} or {\sl so(2,4)} Lie algebras with generators $t_{IJ}$ with $I,J=1,\dots,6$. We define a topological gauge theory on a five dimensional manifold ${\cal M}_5$ by the integral of the CS 5-form  ${\cal L}_5$
\ba\label{CS} S = \alpha \int_{{\cal M}_5} {\cal L}_5&=&\alpha\int_{{\cal M}_5}{\rm tr} [{\bf F}\wedge{\bf F}\wedge{\bf A}-{\textstyle\frac{1}{2}}{\bf F}\wedge{\bf A}\wedge{\bf A}\cr&&\qquad\qquad +{\textstyle\frac{1}{10}}{\bf A}\wedge{\bf A}\wedge{\bf A}\wedge{\bf A}\wedge{\bf A}],\quad\ea
where $\alpha$ is an arbitrary constant and $\bf F$ is the curvature 2-form ${\bf F}={\rm d}{\bf A}+{\bf A}\wedge{\bf A}$. The CS 5-form satisfies
\be d{\cal L}_5 = {\rm tr}[{\bf F}\wedge{\bf F}\wedge{\bf F}]=\epsilon_{IJKLMN}F^{IJ}\wedge F^{KL}\wedge F^{MN},\ee
where $\epsilon$ is the group invariant tensor. The action, by construction, is invariant (up to a boundary term) under the following  gauge transformations with parameters $\Lambda^{IJ}$
\be A^{IJ}\rightarrow A^{IJ}+d\Lambda^{IJ}+A^I\,_K\Lambda^{KJ}-A^J\,_K\Lambda^{KI}.\ee
The equations of motion are computed as
\be \epsilon_{IJKLMN} F^{KL}\wedge F^{MN}=0.\ee
If the five dimensional manifold has boundaries, boundary terms must be vanishing too.We note that this gauge theory has 13 propagating degrees of freedom.

\paragraph*{Group decomposition:}
For later application, we decompose of the gauge connection in {\sl SO(1,4)} (or {\sl SO(2,3)}) covariant form which is represented as
\be A^{IJ} =
\bigg[
\begin{array}{cc}
\omega^{AB} &\ \ e^A\\
-e^B  &\ 0   
\end{array}
\bigg],\ee
where $A,B=1,2,\dots,5$. Then, the curvature is
\be F^{IJ} =
\bigg[
\begin{array}{cc}
R^{AB} \mp e^A\wedge e^B&\ \ De^A\\
-De^B  &\ 0   
\end{array}
\bigg],
\ee
where the upper sign is for the dS and the lower sign is for the adS group. Moreover, the covariant derivative is
\be (De)^A={\rm d}e^A+\omega^A\,_Be^B.\ee
We write the action in $SO(1,4)( {\rm or}\ SO(2,3))$ notation
\ba\label{reduced-CS} {\cal L}_5\!=\! 3\alpha\epsilon_{ABCDE} \big[&&R^{AB}\!\wedge\! R^{CD}\!\wedge\! e^E
\mp{\textstyle\frac{2}{3}}R^{AB}\!\wedge\! e^C\!\wedge\! e^D\!\wedge\! e^E\cr
&&\qquad\qquad+{\textstyle\frac{1}{5}}e^A\!\wedge\! e^B\!\wedge\! e^C\!\wedge\! e^D\!\wedge\! e^E\big].\ea

\paragraph*{Interpretation as a theory of gravity}
There is no dimensional parameter in the action \eqref{CS} or \eqref{reduced-CS}.  After splitting the gauge connection $A$ to $\omega$ and $e$ we introduce a scale $l$ through $e\rightarrow l^{-1}e$ redefinition so that the connection $e$ is dimensionless. Then, we identify $\omega$ as the spin-connection and $e$ as funfbein. In the space of invertible funfbeins, we can define metric structure and interpret \eqref{reduced-CS} as special case of Lovelock action for gravity in five dimensions with the Planck mass $M_5=(\mp4\alpha)^{1/3}l^{-1}$. 
Through taking limits $l\rightarrow\infty$ or $l\rightarrow0$ and properly rescaling $\alpha$ we respectively find the Gauss-Bonnet gravity (with contracted {\sl ISO(1,4)} symmetry) and non-dynamical cosmological constant term.  However, there is no limit where we find soley the Einstein-Cartan term. 


\paragraph*{Further group decomposition:} We split the connection 1-form in terms of {\sl SO(1,3)} representations as
\be \omega^{AB} =
\bigg[
\begin{array}{cc}
\omega^{ab} &\ \ f^a\\
-f^b  &\ 0   
\end{array}
\bigg],\ee
where now $a,b=1,2,3,4$. Thus, the curvature will be 
\be R^{AB} =
\bigg[
\begin{array}{cc}
R^{ab} - f^a\wedge f^b&\ \ Df^a\\
-Df^b  &\ 0   
\end{array}
\bigg],
\ee
where now the covariant derivative is defined as
\be (Df)^a={\rm d}f^a+\omega^a\,_bf^b.\ee The {\sl SO(1,4)} vector is also decomposed 
\be e^A = 
\bigg[
\begin{array}{c}
e^a \\ \tilde e
\end{array}
\bigg].\ee
Therefore, the CS 5-form is represented as follows
\ba {\cal L}_5 =3\alpha\epsilon_{abcd}\big[R&&^{ab}\wedge R^{cd}\wedge \tilde e\cr
-&&2R^{ab}\wedge f^c\wedge f^d\wedge \tilde e\cr
+&&2R^{ab}\wedge e^c\wedge Df^d \cr
\mp&&2R^{ab}\wedge e^c\wedge e^d\wedge \tilde e\cr
+&&f^a\wedge f^b\wedge f^c\wedge f^d\wedge \tilde e\cr
-&&2e^a\wedge f^b\wedge f^c\wedge  Df^d\cr
\pm&&2e^a\wedge e^b\wedge f^c\wedge f^d\wedge \tilde e\cr
\mp&&{\textstyle\frac{2}{3}}e^a\wedge e^b\wedge e^c\wedge Df^d\cr
+&&e^a\wedge e^b\wedge e^c\wedge e^d\wedge \tilde e
\big].\ea
The above action is invariance under the following {\sl SO(1,5)} (or {\sl SO(2,4)})  gauge transformations
\begin{subequations}\begin{align}
\label{gf-1}\delta\omega^{ab} &= (D\lambda)^{ab}-f^{[a}\kappa^{b]}\mp e^{[a}\epsilon^{b]},\\
\label{gf-2}\delta f^a &= (D\kappa)^{a}+f^b\lambda_b\,^a\mp e^a\varepsilon\pm \tilde e \epsilon^a,\\
\label{gf-3}\delta e^a &=(D\epsilon)^{a}+e^b\lambda_b\,^a+f^a\varepsilon-\tilde e\kappa^a,\\
\label{gf-4}\delta \tilde e &= {\rm d}\varepsilon-f^a\kappa_{a}+e^a\epsilon_{a},
\end{align}\end{subequations}
where $\Lambda^{ab}=\lambda^{ab}$ are parameters of {\sl SO(1,3)} transformations and we define $\Lambda^{a6}=\epsilon^a$, $\Lambda^{a5}=\kappa^a$ and $\Lambda^{56}=\varepsilon$.

In the final step, we split the fifth dimension from the other four and express the dynamical fields as
\begin{subequations}\begin{align}
\omega^{ab}&=\omega^{ab}_\mu(x,y){\rm d}x^\mu+\omega^{ab}_y(x,y){\rm d}y,\\
f^{a}&=f^{a}_\mu(x,y){\rm d}x^\mu+f^{a}_y(x,y){\rm d}y,\\
e^{a}&=e^{a}_\mu(x,y){\rm d}x^\mu+e^{a}_y(x,y){\rm d}y,\\
\tilde e&=\tilde e_\mu(x,y){\rm d}x^\mu+\tilde e_y(x,y){\rm d}y.
\end{align}\end{subequations}
Then, the action is rewritten \ba {\cal L}_5=3\alpha&&\epsilon_{abcd}\Big[
R^{ab}\wedge R^{cd}\wedge \tilde e_y+6R^{ab}\wedge \tilde e\wedge R_y^{cd} \cr
-&&2R^{ab}\wedge f^c\wedge f^d\wedge \tilde e_y+4R^{ab}\wedge f^c\wedge \tilde e\wedge f_y^d\cr 
-&&2f^a\wedge f^b\wedge \tilde e\wedge R_y^{cd}+2R^{ab}\wedge e^c\wedge (Df)^d_y\cr
-&&6R^{ab}\wedge (Df)^c\wedge e_y^d+2 e^a\wedge (Df)^b\wedge R_y^{cd}\cr
\mp&&2R^{ab}\wedge e^c\wedge e^d\wedge \tilde e_y\pm 4R^{ab}\wedge e^c\wedge \tilde e\wedge e_y^d\cr
\mp&&2e^a\wedge e^b\wedge \tilde e\wedge R_y^{cd}+f^a\wedge f^b\wedge f^c\wedge f^d\wedge \tilde e_y\cr 
-&&4f^a\wedge f^b\wedge f^c\wedge \tilde e\wedge f_y^d- 2e^a\wedge f^b\wedge f^c\wedge(Df)^d_y\cr +&&4e^a\wedge f^b\wedge (Df)^c\wedge f_y^d+2f^a\wedge f^b\wedge (Df)^c\wedge e_y^d\cr
\pm&&2e^a\wedge e^b\wedge f^c\wedge f^d\wedge \tilde e_y\mp 4 e^a\wedge e^b\wedge f^c\wedge \tilde e\wedge f_y^d\cr
\mp&&4 e^a\wedge f^b\wedge f^c\wedge \tilde e\wedge e_y^d\mp{\textstyle\frac{2}{3}}e^a\wedge e^b\wedge e^c\wedge (Df)^d_y \cr
\pm&&2e^a\wedge e^b\wedge (Df)^c\wedge e_y^d + e^a\wedge e^b\wedge e^c\wedge e^d\wedge \tilde e_y \cr -&&4e^a\wedge e^b\wedge e^c\wedge \tilde e\wedge e_y^d\Big],\ea
where the subscript $y$ means that the differential forms have one leg along the $5^{th}$ direction. In particular 
\ba R_y&=&R^{ab}_{y\mu}{\rm d}x^\mu\wedge {\rm d}y ,\cr
(Df)_y&=&D_yf_\mu^a{\rm d}x^\mu\wedge{\rm d}y+D_\mu f_y^a{\rm d}y\wedge{\rm d}x^\mu.\ea
The equations of motion of $\tilde e_y,e^a_y,f^a_y,\omega^{ab}_y, \tilde e_\mu,e^a_\mu,f^a_\mu$ and $\omega^{ab}_\mu$ are computed respectively as follows
\ba 0&=&\epsilon\!\cdot\! (R-f\wedge f\mp e\wedge e)\wedge (R-f\wedge f\mp e\wedge e),\\ 0&=&\epsilon\!\cdot\! (R-f\wedge f\mp e\wedge e)\wedge(Df\pm2e\wedge \tilde e),\\
0&=&\epsilon\!\cdot\! (R-f\wedge f\mp e\wedge e)\wedge(De+2f\wedge \tilde e),\\
0&=&\epsilon\!\cdot\! [(d\tilde e-f\wedge e)\wedge(R-f\wedge f\mp e\wedge e)\cr&&\quad -2(Df\pm2e\wedge \tilde e)\wedge(De+2f\wedge \tilde e)],\\
0&=&\epsilon\!\cdot\! (R-f\wedge f\mp e\wedge e)\wedge (R_y\!-2f\wedge f_y\mp 2e\wedge e_y),\\
0&=&\epsilon\!\cdot\! (R-f\wedge f\mp e\wedge e)\wedge((Df)_y\mp2e\wedge \tilde e_y\mp2e_y\wedge \tilde e)\cr &+&\epsilon\!\cdot\! (R_y-2f\wedge f_y\mp 2e\wedge e_y)\wedge(Df\pm2e\wedge \tilde e),\ \\
0&=&\epsilon\!\cdot\! (R-f\wedge f\mp e\wedge e)\wedge((De)_y-2f\wedge \tilde e_y-2f_y\wedge \tilde e)\cr &+&\epsilon\!\cdot\! (R_y-2f\wedge f_y\mp 2e\wedge e_y)\wedge(De+2f\wedge \tilde e),\\
0&=&\epsilon\!\cdot\! [(d\tilde e_y-f\wedge e_y+f_y\wedge e)\wedge(R-f\wedge f\mp e\wedge e)\\&&
\quad +(d\tilde e-f\wedge e)\wedge(R_y-2f\wedge f_y\mp 2e\wedge e_y)\cr
&&\quad+2((Df)_y\pm2e\wedge \tilde e_y\mp2e_y\wedge \tilde e)\wedge(De+2f\wedge \tilde e)\cr&&
\quad-2((De)_y+2f\wedge \tilde e_y-2f_y\wedge \tilde e)\wedge (Df\pm2e\wedge \tilde e)].\nn
\ea

\paragraph*{Partial gauge fixing:}The action is invariant under fifteen gauge transformations \eqref{gf-1} through \eqref{gf-4}. The parameters of transformations ($\Lambda(x,y)=\{\lambda^{ab}, \kappa^a, \epsilon^a, \varepsilon\}$) can be expanded as follows
\be \Lambda(x,y)=\lambda(x)+{\textstyle\sum}_i\lambda_i(x)\varphi_i(y),\ee
where $\varphi_i$ are complete orthonormal functions in one dimensions. 
We use some part of the gauge transformations to fix the fifth components of the gauge fields everywhere in the bulk. However, for later application, we require that the gauge fixed action is still invariant under {\sl SO(1,3)} gauge symmetries. Thus, we fix a gauge as
\ba e_y^a(x,y) &=&0,\\ \label{g-2}f^a_y(x,y)&=&0,\\ \omega^{ab}_y(x,y)&=&0.\ea
Moreover, $\tilde e_y$ can be fixed to by the remaining gauge transformation \eqref{gf-4}. It is in scalar representation of 4-dimensional Lorentz transformations. Thus, it can be fixed to either zero or a non-zero value. We are interested in non-zero value as it break the scale symmetry of the theory by introducing a dimensionful constant $l$ as 
\be\label{e-tilde}\tilde e_y(x,y)=l^{-1}.\ee
In this gauge, the 5-dimensional Lagrangian is given by
\ba\label{gf-action} {\cal L}_5=3\alpha \epsilon_{abcd} \big[
&&l^{-1} R^{ab}\wedge R^{cd}\cr
&&-2 l^{-1}R^{ab}\wedge f^c\wedge f^d\cr 
&&+2R^{ab}\wedge e^c\wedge \partial_yf^d\cr
&&+2 e^a\wedge Df^b\wedge \partial_y\omega^{cd} \cr
&&+2R^{ab}\wedge\tilde e\wedge\partial_y\omega^{cd}\cr
&&-2f^a\wedge f^b\wedge\tilde e\wedge\partial_y\omega^{cd}\cr
&&\mp2e^a\wedge e^b\wedge\tilde e\wedge\partial_y\omega^{cd}\cr
&&\mp2 l^{-1}R^{ab}\wedge e^c\wedge e^d\cr
&&+l^{-1} f^a\wedge f^b\wedge f^c\wedge f^d\cr 
&&-2e^a\wedge f^b\wedge f^c\wedge \partial_yf^d\cr
&&\pm2 l^{-1}e^a\wedge e^b\wedge f^c\wedge f^d\cr
&&\mp{\textstyle\frac{2}{3}}e^a\wedge e^b\wedge e^c\wedge\partial_y f^d\cr 
&&+l^{-1} e^a\wedge e^b\wedge e^c\wedge e^d\big]\wedge{\rm d}y.
\ea
Needless to say, in a gauge with $\tilde e_y=0$ ($l\rightarrow 0$) we find a different theory that we abandon to study in this paper. 

\subsection*{III. Symmetry breaking by boundary conditions}
We assume that the 5-dimensional manifold has boundaries along the fifth direction. The bulk equations of motions for respectively $\tilde e$, $e^a$, $f^a$ and $\omega^{ab}$ are
\ba\label{eom-1} 0&=&\epsilon_{abcd}(R^{ab}\mp e^a\wedge e^b-f^a\wedge f^b)\wedge \partial_y\omega^{cd},
\\
0&=&\epsilon_{abcd}(R^{bc}\mp e^b\wedge e^c-f^b\wedge f^c)\wedge (\mp 2e^d+\partial_yf^d)\qquad\cr &+&\epsilon_{abcd}(Df)^b\wedge \partial_y\omega^{cd},\\
0&=&\epsilon_{abcd}(R^{bc}\mp e^b\wedge e^c-f^b\wedge f^c)\wedge (- 2f^d+\partial_ye^d)\cr &+&\epsilon_{abcd}(De)^b\wedge \partial_y\omega^{cd},\\
\label{eom-4}0&=&\pm \epsilon_{abcd}e^c\wedge (De)^d+\epsilon_{abcd}f^c\wedge (Df)^d\cr
&-&{\textstyle\frac{1}{2}}\epsilon_{abcd}(De)^c\wedge\partial_yf^d-{\textstyle\frac{1}{2}}\epsilon_{abcd}(Df)^c\wedge\partial_ye^d.
\ea
A simple familiar class of solutions are given by
\ba 
(De)^a&=&0,\\
(Df)^a&=&0,\\
R^{ab}\mp e^a\wedge e^b-f^a\wedge f^b &=&0.
\ea

Moreover, the variation of the action gives the boundary terms that must be vanishing
\be\label{boundary-term} \int_{{\cal M}_4}\int_{y=0}^{y=L}\partial_y\Big[\frac{\partial {\cal L}}{\partial(\partial_yf^a)}\delta f^a+\frac{\partial {\cal L}}{\partial(\partial_y\omega^{ab})}\delta \omega^{ab}\Big]=0.\ee
In the following, we assume that each term in \eqref{boundary-term} is vanishing independently on the endpoints which yield the following conditions
\ba\label{BC}  0&\!=\!&\epsilon_{abcd}e^a\wedge(R^{bc}\mp \sfrac{1}{3}e^b\wedge e^c-f^b\wedge f^c )\wedge \delta f^d\big{|}_{y=0}^{y=L},\\
\label{constraint-2}0&\!=\!&\epsilon_{abcd}(e^a\wedge Df^b\!+\!R^{ab}\mp e^a\wedge e^b\!-\!f^a\wedge f^b)\!\wedge\delta\omega^{cd}\big{|}_{y=0}^{y=L}.\ \nn\\
\ea
There are variety of choices for boundary conditions so that they satisfy the above requirements. The simplest ones are that either the variations of the fields $f^a$ and $\omega^{ab}$ vanish at both endpoints
\ba\label{BC-1} \delta f^d\big{|}_{y=0\ {\rm and}\ \pi L}&=&0 ,\\ \delta\omega^{cd}\big{|}_{y=0\ {\rm and}\ \pi L}&=&0, \ea
or so do their coefficients 
\ba 
R^{ab}\mp \sfrac{1}{3}e^a\wedge e^b-f^a\wedge f^b\big{|}_{y=0\ {\rm and}\ \pi L}  &=&0,\ \\
\label{condition-2}e^a\wedge Df^b\!+\!R^{ab}\mp e^a\wedge e^b\!-\!f^a\wedge f^b\big{|}_{y=0\ {\rm and}\ \pi L}&=&0.\ \ea
We limit ourselves to these simplest conditions. 

The general solution to \eqref{BC-1} is
\ba f^a_\mu(x,y=0)=\zeta^a_\mu\ &,&\ f^a_\mu(x,y=L)=\tilde \zeta^a_\mu,\\
\omega^{ab}_\mu(x,y=0)=\zeta^{ab}_\mu\ &,&\ \omega^{ab}_\mu(x,y=L)=\tilde \zeta^{ab}_\mu,
\ea
where $\zeta$ and $\tilde\zeta$  are constants vectors/tensors with mass dimension one. We note that for $\zeta=\tilde \zeta=0$ all gauge symmetries are preserved in lower dimensions. 

Generic choices of constants ($\zeta,\tilde \zeta\neq 0$) break all or some of the gauge symmetries. However, there is a particular non-trivial choice that a global diagonal $SO(1,3)$ symmetry is preserved. We recall that the Lagrangian \eqref{gf-action} is invariant under $SO(1,3)$ of internal gauge transformation and another $SO(1,3)\times U(1)\subset$ {\sl Diff(5)} which originated from the topological invariance of the CS action. Boundary conditions at the endpoints that satisfies the first equation in \eqref{BC} can be chosen so that a diagonal $SO(1,3)$ is preserved as follows
\ba\label{cond-1}  f_\mu^a(x,y=0)&=&\zeta\delta_\mu^a\quad {\rm and}\cr
f_\mu^a(x,y=\pi L)&=&\tilde \zeta\delta_\mu^a,\ea
where $\zeta$ is a dimensional constant of either sign. Equivalently, it can be represented in terms of differential  forms
\be f^a(x,y=0)= \zeta{\bf 1}^a\quad  {\rm and}\quad f^a(x,y=\pi L)=\tilde \zeta {\bf 1}^a.\ee
where ${\bf 1}^a$ are constant 1-forms. We emphasis that $\zeta$ and $\tilde\zeta$ are order parameters of symmetry breaking.

The second condition in \eqref{constraint-2} can be satisfied given that
\be\label{cond-2} \omega^{ab}(x,y=L)=\omega^{ab}(x,y=0)=0.\ee
Alternatively, one can choose to satisfy \eqref{constraint-2} by admitting to \eqref{condition-2} and demands
\ba (Df)^a\big{|}_{y=0\ {\rm and}\ L}&=&0, \\ R^{ab}\mp e^a\wedge e^b-f^a\wedge f^b\big{|}_{y=0\ {\rm and}\ L}&=&0.\ea
The first constrain implies that the Lorentz vector $f^a$ is covariantly constant at endpoints.

In the following, we choose \eqref{cond-1} and \eqref{cond-2}. Moreover, there is enough residual gauge freedom to require that 
\be\label{more-gf} \partial_\mu f_\mu^a(x,y)=0.\ee
In fact, we applied gauge transformation \eqref{gf-2} to fix the fifth component of $f^a$ as in \eqref{g-2}. However, we can add to the gauge parameters in \eqref{gf-2} some functions of  4-dimensional coordinates $x^\mu$ so that the gauge in \eqref{g-2} is preserved. This freedom is enough to satisfy \eqref{more-gf}. For $\zeta\neq0$ there are three choices for $\tilde \zeta$ so that $SO(1,3)$ is preserved, namely $\tilde \zeta=0$, $\tilde \zeta=-\zeta$ and $\tilde \zeta=\zeta$. 
 The KK decomposition of $f^a_\mu$ fields for three choices are
\ba\label{f-expansion} f^a_\mu(x,y)&=&\zeta\delta_\mu^a\cos(c L^{-1}y),\ea
where $c$ is either of $\sfrac{1}{2}, 1, 2$ for different choices of $\tilde \zeta$. Without loss of generality, we choose $c=1$ in the following analysis. 

The KK expansion of $e^a$ and $\omega^{ab}$ fields are periodic on the interval as written as
\ba
\label{e-expansion}e^a_\mu(x,y)&=&e^a_{\mu}(x)+\textstyle\sum_{n=1} e^{a,n}_{\mu}(x)\cos(n L^{-1}y),\\
\label{omega-expansion}\omega^{ab}_\mu(x,y)&=&\omega^{ab}_{\mu}(x)+\textstyle\sum_{n=1}\omega^{ab,n}_{\mu}(x)\cos(n L^{-1}y).
\ea

Finally, we note that when we assign a coordinate system to the 5-dimensional manifold, we can fix four coordinates (by infinite dimensional diffeomorphisms) so that $\tilde e_\mu{\rm d}x^\mu=0$. The other freedom along the fifth coordinate solely rescales the gauge choice in \eqref{e-tilde}.  Therefore using all the above freedom, we are allowed to greatly simplify the action and (ignoring a shift by real numbers) as 
\ba\label{relevant-action} {\cal L}_5=3\alpha \epsilon_{abcd} \big[
&&\mp2 l^{-1}R^{ab}\wedge e^c\wedge e^d\cr
&&-2e^a\wedge f^b\wedge f^c\wedge \partial_yf^d\cr
&&\pm2 l^{-1}e^a\wedge e^b\wedge f^c\wedge f^d\cr
&&\mp{\textstyle\frac{2}{3}}e^a\wedge e^b\wedge e^c\wedge\partial_y f^d\cr 
&&+l^{-1} e^a\wedge e^b\wedge e^c\wedge e^d\big]\wedge{\rm d}y.
\ea

\subsection*{Dimensional reduction in 4 dimensions}
Now, we are ready compactify the simplified 5-dimensional CS theory to get an effective 4-dimensional theory. The fifth dimension is compactified on a line segment of length $L$ so that $\int {\rm d}y=L$. 
Basically, the kinetic terms and interactions among different KK modes are obtained by substituting fields expansion into the action and integrating over the extra dimension. 
\paragraph*{Interpretation as a 4-dimensional theory of gravity}
In order to interpret the reduced theory as a gravitational theory, we take $e^a_\mu$ as invertible vierbeins and rescale to make them dimensionless 
\be e^a\rightarrow l^{-1}e^a.\ee
Consequently, we can endow the manifold with a metric defined by $g_{\mu\nu}=e^a_\mu e^b_\nu\eta_{ab}$ and the general covariance is descended from the topological invariance. 

Then apparently, the first term in \eqref{relevant-action} yields the Einstein-Hilbert action which gives the kinetic terms, mass terms and some part of interactions among KK states. The 4-dimensional Planck mass is identified as
\be\label{planck-mass} m_{\rm Pl}^2=\mp24\alpha l^{-3}L.\ee 
From above, we can identify the scale $l$ we the 5-dimensional Planck length and in order for geometry to make sense we must have $L< l$ and so $\mpl>l^{-1}$. We recall that the free parameter $\alpha$ can be of either sign. 

Before we continue, we point out that the mass of level $n$ KK modes are given by 
\be m_{\rm KK}^{(n)}=nL^{-1}.\ee
The validity of the lower-dimensional effective field theory requires that $m_{\rm KK}^{(n)}<l^{-1}$. Consequently, the 4-dimensional theory includes KK modes up to level $n$ given by $n<Ll^{-1}$.

Now we work out the potential terms by substituting the fields expansion \eqref{f-expansion} and \eqref{e-expansion} and integrate over the extra dimension. On zero modes we find
\ba {\cal V}\supset&&-(3\alpha l^{-5}L)\epsilon_{abcd}e^a\wedge e^b\wedge e^c\wedge e^d\cr
&&\mp(4\alpha \zeta l^{-3})\epsilon_{abcd}e^a\wedge e^b\wedge e^c\wedge {\bf 1}^d\cr
&&\mp(3\alpha \zeta^2l^{-3}L)\epsilon_{abcd}e^a\wedge e^b\wedge {\bf 1}^c\wedge{\bf 1}^d\cr
&&-(4\alpha \zeta^3l^{-1})\epsilon_{abcd}e^a\wedge {\bf 1}^b\wedge {\bf 1}^c\wedge {\bf 1}^d
.\ea
We read the cosmological constant from the first term as 
\be \Lambda=-3\alpha l^{-5}L=\pm{\textstyle\frac{1}{8}} l^{-2}m_{\rm Pl}^2.\ee
Next, we compare this with the 2-parameter family of the dRGT potential terms \cite{Hinterbichler:2012cn}
\ba\label{dRGT-potential} {\cal V}_{\rm dRGT} &=& \sfrac{1}{24}(m^2\mpl^2)b_0\epsilon^{abcd}e^a\wedge e^b\wedge e^c\wedge e^d\cr
&+&\sfrac{1}{6}(m^2\mpl^2)b_1\epsilon^{abcd}e^a\wedge e^b\wedge e^c\wedge {\bf 1}^d\cr
&+&\sfrac{1}{4}(m^2\mpl^2)b_2\epsilon^{abcd}e^a\wedge e^b\wedge {\bf 1}^c\wedge {\bf 1}^d\cr
&+&\sfrac{1}{6}(m^2\mpl^2)b_3\epsilon^{abcd}e^a\wedge {\bf 1}^b\wedge {\bf 1}^c\wedge {\bf 1}^d,
\ea 
where $m$ is the graviton mass. Therefore, we determine the free parameters in the dRGT potential in terms of the fundamental parameters induced by the higher-dimensional theory 
\ba b_0m^{2}&=&\pm3l^{-2}\cr
 b_1m^{2}&=&\zeta L^{-1},\cr 
 b_2m^{2}&=&{\textstyle\frac{1}{2}}\zeta^2,\cr 
 b_3m^{2}&=&\pm\zeta^3 l^{2}L^{-1}.\ea 
The graviton mass $m$ is read through the condition $b_1+2b_2+b_3 = 1$. Then , we find that
\be\label{graviton-mass} m^2=\zeta L^{-1}\big[1+\zeta L\pm(\zeta l)^2\big]\approx \zeta L^{-1}(1+\zeta L),\ee
where in the last step we made the natural assumption that the scale of symmetry breaking is less than the fundamental scale of the gravitational theory in 5-dimensions {\it i.e.} $\zeta\ll l^{-1}$. Moreover comparing the scale of symmetry breaking with the interval length, we find that there are two possibilities as follows
\ba m^2&\approx& \zeta^2\qquad\quad\ {\rm for}\qquad \zeta\gg L^{-1},\\
m^2&\approx&  \zeta L^{-1}\qquad {\rm for}\qquad \zeta\ll L^{-1},\ea
Therefore, the scales assume two possible hierarchies
\ba  \zeta< m< L^{-1}<l^{-1}<m_{\rm Pl},\\
 L^{-1}<\zeta= m<l^{-1}<m_{\rm Pl}.\ea
In the first case, the massive graviton in the dRGT potential is lighter than all of the KK gravitons. In the second case, it is in the middle of the spectrum.

Furthermore, in the dRGT theory in order that flat spacetime is a solution one demands $b_0+3b_1+3b_2+b_3 = 0$. It implies that 
\be [3+(\zeta l)^2][2(\zeta l)^2\pm3(\zeta L)]\mp3\zeta L=0.\ee
In the limit $\zeta \ll l^{-1}$ it has a solution for $\zeta\ll L^{-1}$ and $l\approx\sqrt 6 L$. Therefore, the consistent hierarchy of scales is given by the first case.

We emphasis that the proposed  theory involves  four  parameters $\alpha, l, L$ and $\zeta$. One combination of parameters fixes the 4-dimensional Planck mass \eqref{planck-mass}, another combination determines the graviton mass \eqref{graviton-mass} and the other two independent parameters (counterparts of $c_3$ and $d_5$ in the dRGT model) give interaction strengths in the potential terms. It gives two-parameter family of the most general ghost free potential terms. The potential terms are constructed top-down from a CS gauge theory and coefficients are determined through symmetry breaking mechanism and compactification scale.  The UV cutoff of the theory $\Lambda_{\rm UV}$ is the 5-dimensional Planck scale $l^{-1}$ and it is parametrically much greater than the UV cutoff of the dRGT theory $\Lambda_3$ as
\be   \Lambda_{\rm UV}\sim l^{-1}\sim (L^{-1}\mpl^2)^{\sfrac{1}{3}}\gg \Lambda_3\sim (\mpl \zeta L^{-1})^{\sfrac{1}{3}}.\ee
Therefore, we say that the 4-dimensional theory of massive spin-2 is UV completed in five dimensions.

In passing we note that different compactification schemes and different choices for the boundary conditions would yield different spin-2 field theories.

\subsection*{Interaction of Kaluza-Klein modes}
In order to find the interactions among different KK modes, we substitute the field expansions into the higher-dimensional action and integrate over the extra dimensions. In this model on top of standard interactions from the Einstein-Hilbert action, we find extra interactions of KK states induced by the dRGT-like terms. Here we compute the non-vanishing couplings starting from the simplified action \eqref{relevant-action}.
The first class of interactions are computed as follows
\ba {\cal V}&\supset& -(6\alpha\zeta^3 l^{-1}L^{-1})\epsilon_{abcd}{\bf 1}^a\wedge {\bf 1}^b\wedge {\bf 1}^c\wedge {\textstyle\sum_n} e^d_n\cr &&\qquad\qquad\times\int \cos^2(L^{-1}y)\sin(L^{-1}y)\cos(nL^{-1}y)\cr
&=&(12\alpha l^{-1}\zeta^3)\epsilon_{abcd}{\textstyle\sum_n}c_n e^a_n\wedge {\bf 1}^b\wedge {\bf 1}^c\wedge {\bf 1}^d, \ea
where $c_n$ are 
\be c_n=\frac{1}{1-\frac{4n^2}{n^2-3}}, \qquad n={\rm 2,4,6,\dots}\ .\ee
The other interactions are given by
\ba {\cal V}&\supset& \mp(6\alpha\zeta^2 l^{-1}\epsilon_{abcd}{\bf 1}^a\wedge {\bf 1}^b\wedge{\textstyle\sum_m}e^c_m\wedge{\textstyle\sum_n}e^d_n\cr &&\qquad\qquad\times\int \cos^2(L^{-1}y)\cos(mL^{-1}y)\cos(nL^{-1}y)\cr
&=&\mp(6\alpha l^{-1}L\zeta^2)\epsilon_{abcd}{\textstyle\sum_{m,n}}c_{mn} e^a_m\wedge e_n^b\wedge {\bf 1}^c\wedge {\bf 1}^d, \ea
where the coefficient $c_{mn}$ are
\be c_{mn}={\textstyle\frac{\pi}{8}}\delta_{m+n,2}+{\textstyle\frac{\pi}{8}}\delta_{m-n,2}+{\textstyle\frac{\pi}{4}}\delta_{m,n}.\ee
The lest set of interactions are found as 
\ba {\cal V}&\supset& \mp(2\alpha\zeta l^{-3}L^{-1})\epsilon_{abcd}{\bf 1}^a\wedge\!\int\! \big[e^b\!+\!{\textstyle\sum_m} e_m^b\cos(mL^{-1}y)\big]\cr &&\ \ \qquad\qquad\qquad\qquad\qquad\wedge\big[e^c\!+\!{\textstyle\sum_n} e_n^c\cos(nL^{-1}y)\big]\cr&&\qquad\qquad\qquad\wedge\big[e^d\!+\!{\textstyle\sum_p} e_p^d\cos(nL^{-1}y)\big] \times\sin(L^{-1}y)\cr
&=&\pm(12\alpha\zeta l^{-3})\epsilon_{abcd}{\textstyle\sum_{m}}c_{m} e^a\wedge e^b\wedge e^c_m\wedge {\bf 1}^d \cr
&&\pm(12\alpha\zeta l^{-3})\epsilon_{abcd}{\textstyle\sum_{m,n}}c_{mn} e^a\wedge e^b_m\wedge e^c_n\wedge {\bf 1}^d \cr
&&\pm(\alpha\zeta l^{-3})\epsilon_{abcd}{\textstyle\sum_{m,n,p}}c_{mnp} e^a_m\wedge e^b_n\wedge e^c_p\wedge {\bf 1}^d, \ea
with the following coefficients
\ba c_m&=&\frac{1}{m^2-1}, \qquad m=2,4,6,\dots,\\
c_{mn}&=&\frac{m^2+n^2-1}{(m^2-n^2)^2-2(m^2+n^2)+1},\Big|^{m,n=2,4,6,\dots}_{m,n=1,3,5,\dots},\ \ \  \\
c_{mnp}&=&\frac{(-1)^{m-n-p}}{m-n-p-1}\!+\!\frac{(-1)^{m+n-p}}{m+n-p+1}\!+\!\frac{(-1)^{m-n+p}}{m-n+p+1}.\nn\\
\ea


\subsection*{IV. Conclusion}
In this paper, we revisited the dRGT interaction terms in a theory of massive gravity in a top-down approach. We built an effective field theory in four dimensions with massive spin-2 excitations and no massless one. We got that by a particular dimensional reduction of CS gauge theory in five dimensions. All the parameters of the dRGT theory is computed in terms of the fundamental and the geometric quantities of the 5-dimensional theory. The extra dimension helped to break all the gauge symmetry by appropriate boundary conditions and thus explained the absence of the massless spin-2 particle. Moreover, it provided the lower dimensional theory with a whole tower of KK states. The exchange of these modes improves the high energy behavior of the scattering amplitudes involving the longitudinal modes. Therefore, the UV cutoff the effective field theory in four dimensions is the Planck scale of the 5-dimensional gravitational theory and, unlike the dRGT theory, it is independent of the graviton mass (namely, the IR parameter). These interesting features are descended from the gauge structure of the higher-dimensional theory which is softly broken in lower dimensions.

\paragraph*{Acknowledgments}
This work is supported by the research deputy of Sharif University of Technology.

\ \\$^*$ Email: {\tt mahdi.torabian@sharif.edu}

\end{document}